\documentclass[debug]{rmaa}

\usepackage{paralist}

\usepackage{psfrag,color}

\usepackage[latin1]{inputenc}

\title{The Present and Future of Planetary Nebula Research. A White Paper by the IAU Planetary Nebula Working Group} 

\author{K. B. Kwitter\altaffilmark{1}, R. H. M{\'e}ndez\altaffilmark{2}, M. Pe\~na\altaffilmark{3}, L. Stanghellini\altaffilmark{4},
R. L. M. Corradi\altaffilmark{5,6},
O. De Marco\altaffilmark{7},
X. Fang\altaffilmark{8,9},
R. B. C. Henry\altaffilmark{10},
 A. I. Karakas\altaffilmark{11},
X.-W. Liu\altaffilmark{8,9},
J. A. L\'opez\altaffilmark{12},
A. Manchado\altaffilmark{5,6},
and Q. A. Parker\altaffilmark{7}}

\altaffiltext{1}{Department of Astronomy, Williams College, 33 Lab Campus Drive, Williamstown, MA 01267, USA}
\altaffiltext{2}{Institute for Astronomy, University of Hawaii, 2680 Woodlawn Drive, Honolulu, HI 96822, USA}
\altaffiltext{3}{Instituto de Astronom{\'\i}a, Universidad Nacional Aut\'onoma de M\'exico, 04510 M\'exico, D.F., M\'exico}
\altaffiltext{4}{National Optical Astronomy Observatory, 950 N. Cherry Ave., Tucson, AZ 85719, USA}
\altaffiltext{5}{Instituto de Astrof{\'\i}sica de Canarias, E-38200 La Laguna, Tenerife, Spain}
\altaffiltext{6}{Departamento de Astrof{\'\i}sica, Universidad de La Laguna, E-38206 La Laguna, Tenerife, Spain}
\altaffiltext{7}{Department of Physics \& Astronomy, Macquarie University, Sydney, NSW 2109, Australia}
\altaffiltext{8}{Department of Astronomy, School of Physics, Peking University, Beijing 100871, China}
\altaffiltext{9}{Kavli Institute for Astronomy and Astrophysics, Peking University, Beijing 100871, China}
\altaffiltext{10}{H.L. Dodge Department of Physics and Astronomy, University of Oklahoma, Norman, OK 73019, USA}
\altaffiltext{11}{Research School of Astronomy \& Astrophysics, Mt. Stromlo Observatory,
Weston Creek ACT 2611, Australia.}
\altaffiltext{12}{Instituto de Astronom{\'\i}a, Universidad Nacional Aut\'onoma de M\'exico, Campus Ensenada, Ensenada, B.C., M\'exico}

\shorttitle{PNe White Paper}

\listofauthors{K. B. Kwitter, R. M\'endez, M. Pe\~na, L. Stanghellini, R. L. M. Corradi, O. De Marco, A. I. Karakas, X.-W. Liu, J. A. L\'opez, A. Manchado, Q. A. Parker,  }

\abstract{}

\abstract{We present a summary of current research on planetary nebulae and their central stars, and related subjects such as atomic processes in ionized nebulae, AGB and post-AGB evolution. Future advances are discussed that will be essential to substantial improvements in our knowledge in the field.}

\addkeyword{ISM: abundances}
\addkeyword{planetary nebulae: general}
\addkeyword{stars: AGB and post-AGB}
\addkeyword{stars: evolution} 
\resumen{}

\begin{document}
\maketitle
\section{Introduction}

The field of planetary nebula (PN) research has continued to mature, incorporating a variety of new and ingenious approaches that leverage the value of PN observations to maximize astrophysical insight. PNe are the gaseous relics of the evolution of low- and intermediate-mass stars, and as such are ubiquitous in the Galaxy and beyond. They are probes of stellar evolution, populations, gas dynamics, dust and molecules. They are also extragalactic probes of metallicity and dynamics in spiral and elliptical galaxies and in the intracluster medium.
Spectra of PNe are characteristic and easy to identify; their emission lines are very strong and can be used to derive C, N, O abundances that in turn characterize the stellar progenitor by comparison of the spectra with the yields from stellar evolution theory.
PNe can evolve in binary systems and their morphology and chemical history reflect this type of evolution. They can survive in the intracluster medium, and are rare probes of this interesting environment. Extragalactic PNe have been studied to disclose a characteristic luminosity function whose high-luminosity cutoff appears invariant (or almost invariant) across PN populations, providing a good standard candle. After decades of searching in a variety of celestial objects, researchers have observed fullerenes in PNe, the first environment of stellar origin where these molecules have been observed.

As this brief list reveals, the modern study of PNe is extremely fruitful, with many connections to adjacent fields of research, including stellar structure and evolution, binary stars, stellar populations, radial metallicity gradients in spiral galaxies and their evolution, as well as galaxy rotation, evolution, merging, and cosmology. This White Paper by the IAU PN Working Group, represents a summary of our science activities, and is an attempt to set the stage for future developments in the field. Its aims are to raise interest and to spark discussion within the international PN community, as well as across other interested communities, and finally, to serve as preparation for the next PN Symposium several years hence.

\section{Perspective on Planetary Nebula Detection at Non-optical Wavelengths}

\subsection{The dawn of true multi-wavelength imaging of PNe: Discovery, refinement and characterisation}
Over the last decade Galactic PN discoveries have entered a golden age due to the emergence of high sensitivity, high resolution narrow-band surveys of the Galactic plane (e.g., \citealt{ppp05,dgi05}). These have been coupled with access to complementary, deep, multi-wavelength surveys across near-IR, mid-IR and radio regimes, in particular from both ground-based and space-based telescopes and have provided additional new powerful diagnostic and discovery capabilities.

The total number of Galactic PNe known is currently $\sim$3500, more than double what it was a decade ago. This is largely thanks to the 
$\sim$1200 significant PN discoveries uncovered by the two MASH\footnote{Macquarie/AAO/Strasbourg H$\alpha$ Planetary Galactic Catalog} surveys \citep{paf06,mpa08} based on scrutiny of the SuperCOSMOS AAO/UKST H$\alpha$ survey of the Southern Galactic plane (SHS). 

Note that significant numbers of true PNe are at the very faint, highly evolved end as shown to exist in the local volume sample of \citet{f08}, but they rapidly become undetectable at distances greater than a few kpc. There are also serious problems with obtaining truly representative samples of PNe across the galaxy due to variable extinction.

It is also clear that a significant population of Galactic PNe must be lurking behind the extensive clouds of gas and dust that obscure large regions of our view across the optical regime. Indeed, it is the extension of previous PN discovery techniques away from the optically dominant  [O~III] PN emission line in un-reddened spectra to the longer wavelength  H$\alpha$ emission line (that can peer at least partially through the dust), that has led to the major discoveries of the previous decade. 

Consequently, as part of the mission to improve the inventory of Galactic PNe, extension of PN identification techniques  to longer, more favourable wavelengths would  be advantageous. This is now possible with the advent of powerful, new, multi-wavelength sky-surveys of high sensitivity and resolution. Such surveys now allow us  to both find and accurately characterise the properties of  PNe across a broader range evolutionary state and across a wider range of the electromagnetic spectrum than ever before, opening up new windows into their physical characteristics and dust properties in particular. Such broader perspectives have also given us the capability to not only uncover more PNe but also to refine identification of the many mimics that still lurk in existing PN catalogues.  The improved identification techniques now available are briefly described below.

\subsubsection{Eliminating non-PN contaminants}
Recently \citet{fp10} tested and developed criteria to more effectively eliminate contaminants using the online availability of new multi-wavelength surveys combined with emission line ratios from follow-up spectroscopy.

Applying these criteria to mid-IR samples of previously known optically detected Galactic PNe seen in GLIMPSE\footnote{Galactic Legacy Infrared Mid-Plane Survey Extraordinaire} and that overlap with the SHS ($|b|\leq$1$^{\circ}$ and from 210$^{\circ}$ (through 360$^{\circ}$) to 40$^{\circ}$ in Galactic longitude) showed that 45\% of previously known pre-MASH PNe in this zone are in fact H~II region contaminants \citep{cpg11}. Independently applying these criteria to the MASH PNe revealed a contaminant fraction of 5\% in the same zone, largely because similar discrimination techniques had already been applied to MASH. Furthermore, the external filaments, extended structures and/or amorphous halos that are seen in the mid-IR associated with apparently discrete emission sources in the optical generally indicate that the object is an H~II region. This is an important mid-IR discriminatory diagnostic for resolved mid-IR sources. 

It has also been recently  demonstrated that PN mid-IR/radio flux ratios and the {\it Spitzer} IRAC\footnote{InfraRed Array Camera} colour indices are robust attributes, invariant among PN types including those in nearby external galaxies of different metallicity such as the Large Magellanic Cloud (LMC) \citep{cpg11}. The median PN mid-IR/radio ratio is 4.7$\pm$1.1 and does not vary significantly with PN evolutionary phase, allowing clear separation from their major H~II region contaminants, regardless of whether they are diffuse or ultra-compact. These mid-infrared colours and radio fluxes are at wavelengths minimally affected by dust obscuration. 

The recent advent of the near-infrared Vista Variables in the Via Lactea (VVV) and UKIRT\footnote{United Kingdom Infrared Telescope} Infrared Deep Sky surveys of the Galactic plane have $\sim$4 mag of improved depth and far better resolution compared to 2MASS (Two Micron All Sky Survey) and these, too, offer improved prospects for PN studies over these wavelengths. The judicious combination of the J,H,Ks bands into pseudo-colour images can offer a powerful visual aid as to nature when taken together with the individual near-infrared band photometry. The importance and prospects of the new mid-IR sky surveys to PN studies are summarised in the following section.

\subsubsection{New mid-IR sky surveys and their application to PN science}
PNe can be quite strong mid-IR emitting objects because of PAH (polycyclic aromatic hydrocarbon) emission, fine structure lines, high excitation mid-IR lines like [O~IV]~25.89$\mu$m, H$_2$ molecular lines (e.g. from the UWISH2\footnote{UKIRT Widefield Infrared Survey for H$_2$} survey) and thermal dust emission within the nebulae and in circumnuclear disks. Such emissions make them a decent prospect for uncovering them as mid-IR sources.

More recently, mid-IR space-telescope images from {\it Spitzer} and now WISE (Wide-field Infrared Survey Explorer) potentially allow detection of very reddened PN which may be invisible optically.  Indeed preliminary recent work  noted 416 compact but resolved ($<1$ arcmin) ring, shell and disk-shaped sources in the Galactic plane in 24$\mu$m {\it Spitzer} MIPSGAL\footnote{Multiband Imaging Photometer for Spitzer Galactic Plane Survey} images \citep{mkf10}. Some of these may well turn out to be strongly reddened, high-excitation PNe with perhaps only a minority being circumstellar nebulae around massive stars. 
 
Newly developed mid-IR PN selection techniques have now been established using the refined photometric colour selections of known PNe as the starting point. Such selection criteria now enable efficient searches for highly obscured, previously unknown PNe present within both the photometric source catalogues of currently available Galactic plane on-line mid-IR sky surveys and the associated image data. Nevertheless the techniques have already demonstrated excellent promise with realistic prospects of being able to recognise high quality PN candidates solely using mid-IR and radio characteristics  (e.g. Parker et al. 2012). This would enable efficient trawling for optically hidden PNe when heavy extinction precludes possibility of securing optical spectra and images, alleviating the traditional reliance on such data to indicate their likely PN nature. One area of recent exploitation is in the use of combined multi-band images of PN in the mid-IR as briefly outlined below.

\subsubsection{The power of mid-IR false colour imagery}
\citet{cpg07} have recently shown that true PNe occur with only three colours in combined IRAC band false colour images: red, orange and violet. Similarly 2MASS J,H,Ks false colours of true PNe are violet or pink/purple when they are detected. With the advent of the VVV near infrared Y,Z,J,H,Ks photometric survey one can also trawl for such mid-IR selected PN candidates afresh. VVV overlaps with the very similar J,H,Ks pass-bands of 2MASS but the depth and resolution are far superior  with the VVV J,H,Ks bands extending $\sim4$ mag deeper than 2MASS. \citet{mnc11} have already used the equivalent of the VVV data for the LMC  to show that many PN can be recognised in these new high quality near-IR data. Furthermore, by constructing spectral energy distributions across a broad wavelength range from all the extant wide-field sky surveys it is possible to discriminate among different astrophysical sources including PNe.  The release of WISE data now enables alternative mid-IR false-colour images to be constructed for previously mid-IR selected sources. The WISE 3.4$\mu$m and 4.6$\mu$m bands are directly equivalent to the first two IRAC bands at 3.6$\mu$m and $4.5\mu$m. The final two IRAC bands at 5.8 and 12$\mu$m  do not have any direct WISE equivalent (with the closest being the WISE 12$\mu$m band) though the WISE 22$\mu$m band is similar to the MIPS\footnote{Multiband Imaging Photometer for Spitzer} 24$\mu$m band. WISE can be used as a substitute for IRAC outside of the GLIMPSE regions with excellent sensitivity but poorer resolution. This has major advantages if it can be shown that the sensitivity and resolution of the WISE mid-IR bands are sufficient to provide the same diagnostic capability in revealing PNe as for IRAC. PN candidates can then be  trawled for across the entire sky using essentially the same selection criteria. Examination of known PNe detected in WISE reveal strong potential in this regard.

\subsubsection{The value of the Mid-IR/radio ratio}
\citet{cg01} offered the ratio of the observed flux near 8$\mu$m to that near 1~GHz as a useful object nature discriminant.  It has the merit of well separating different types of H~II regions from PNe if the source is detected in both the mid-IR and radio continuum. The median ratio is 25$\pm$5 for diffuse and compact  regions, and 42$\pm$5 for ultra-and hyper-compact HII regions. For PNe the ratio is 4.7$\pm$1.1.  

\subsection{Summary}
The potential of the available mid-IR survey data from GLIMPSE and WISE as a tool to uncover PN candidates that would be hard or impossible to locate optically is clear. The motivation is to develop mid-IR PN candidate selection techniques that can be used to uncover the significant numbers of Galactic PNe which are believed to be hidden behind extensive curtains of dust. Recent work has also shown that examination of false-colour images of mid-IR selected PN candidates is of high diagnostic value as it enables not just the actual mid-IR photometry to be used for confirmed candidates but crucially the environmental context of sources to also be evaluated.  Ultimately the mid-IR colour-colour techniques recently developed can be applied to  the all sky coverage offered by WISE. In this way mid-IR PNe selected candidates can be compiled not only across the entire area covered by the SHS, IPHAS\footnote{The Isaac Newton Telescope Wide Field Camera Photometric H$\alpha$ Survey}, VPHAS+\footnote{The VLT Survey Telescope/OmegaCam Photometric H$\alpha$ Survey of the Southern Galactic Plane and Bulge} and the VVV but also to higher latitudes where there is no narrow-band coverage. 

\section{Evolutionary Models}
Progenitor masses of PNe are between $\approx 0.8M_\odot$ to
8$M_\odot$, that is, stars heavy enough to ignite helium but avoid core
collapse supernova.  The stellar evolutionary codes used to model 
PN progenitors need prescriptions for dealing with highly uncertain 
phenomena such as convection, mass loss, rotation, and
magnetic fields. Other required input physics include stellar opacities,
thermonuclear reaction rates, and equations of states which 
introduce additional, although smaller, uncertainties. 

\subsection{Convection}
The inclusion of a convective model and a treatment for dealing with convective
boundaries are essential for realistic stellar evolution modeling.
Convection is still the biggest uncertainty in 
evolutionary calculations, a situation that has persisted for more
than 30 years. Only recently has it become possible to
model the convection that occurs in low-mass stars using
hydrodynamical simulations. Most stellar evolution codes use
the mixing-length theory of convection, which
depends on a number of free and uncertain parameters, notably the
mixing-length parameter, $\alpha$. This parameter is usually set by requiring
that a 1$M_\odot$ model of solar composition matches the solar radius.
There are other observables that are now being used to constrain
$\alpha$, e.g., giant branch temperatures for stars in clusters 
\citep{lebzelter07,kamath12}. 
Either way, once $\alpha$ is set it is left constant for all masses 
and evolutionary stages, which is likely to be incorrect.

The persistent issue with convection means that we do not have a good
 description for the third dredge-up \citep[TDU, e.g.,][]{frost96}.
Understanding the TDU is important because it determines 
the chemical enrichment from AGB (asymptotic giant branch) stars into the interstellar
medium. Furthermore, the TDU efficiency  also helps set the
final H-exhausted core mass at the tip of the AGB, which is
important for the initial--final mass relation. 
Stellar evolution models of low-mass AGB stars with near-solar
metallicities still do not obtain efficient TDU without the inclusion of 
convective overshoot \citep[e.g.,][]{cristallo09,karakas10b}.
While it is now standard practice to include  convective overshoot, the
main improvements have come using AGB stars in clusters as a 
constraint on mixing in 
stellar evolution models \citep{lebzelter08, lederer09b, kamath12}.

\subsection{Other theoretical issues}
Observational data have revealed that standard stellar
evolutionary models of low- and intermediate-mass stars (LIMS) are missing key
physics that lead to surface abundance changes during the first and
asymptotic giant branches (e.g., lower $^{12}$C/$^{13}$C ratios
in giant stars than predicted by theory). This extra mixing may
be driven by rotation and magnetic fields
\citep{nordhaus08,lagarde12}, although other phenomena such as
 thermohaline mixing \citep{eggleton08,charbonnel07,stancliffe07b},
and gravity waves \citep{denissenkov03a,talon08} have been suggested.
All of these phenomena occur in stars at some level, so the question is what impact
they have on the stellar interior as a function of time.
Recent  efforts are mostly the result of hydrodynamical simulations
\citep{stancliffe11,herwig11,viallet12}, simply because these 
phenomena are inherently 3D in nature. Future work
should concentrate on higher resolution simulations that are evolved
for longer. However, this requires large supercomputer resources.
3D stellar evolution codes such as {\it DJEHUTY} are fully explicit 
hydrodynamics codes, and hence not suited to evolutionary calculations 
except on the shortest timescales \citep{eggleton08}. Thus 
stellar evolutionary sequences will still be done in 1D, but
improvements in the 1D codes' physics are possible using
the 3D simulations as a guide; this is starting to happen
\citep{arnett11}.  Improvements to our understanding of the TDU
are also likely to come from this area.

Mass loss is now routinely included although there is no agreement
which prescription should be used
during the AGB.  While the prescriptions used in AGB model calculations
today are not new \citep[e.g.,][]{vw93,blocker95a}, {\it Spitzer} has delivered new
insights into mass loss and dust production from LIMS. These are helping to improve 
our understanding of these uncertain processes.  Mass loss determines the
final core mass at the tip of the AGB, and the circumstellar material
around AGB and post-AGB stars. The final ejection of the envelope
is difficult to model in theoretical calculations, and convergence
difficulties often set in before all of the envelope is lost. There
may be some real physics behind this, as explored by \citet{lau12}.

One recent improvement to red giant evolutionary sequences 
has come from the inclusion of low-temperature molecular opacity
tables that follow the surface composition of the star. Low-mass stars are
known to become C-rich during the AGB, and intermediate-mass 
AGB stars with hot bottom burning will become N-rich. These
enrichments lead to highly non-solar C/O and N/C ratios.
The inclusion of molecular opacities following the surface
composition leads to increases in the stellar radius, cooling the
outer layers and leading to stronger mass loss \citep{marigo02}.
New tables have been published in recent
years \citep{lederer09,marigo09}, where the {\it {\AE}SOPUS} tables of 
\citet{marigo09} are the most versatile as they are 
available for a wide range of
initial compositions and can be downloaded from the web.
Recent full AGB simulations 
including some treatment of molecular opacities includes 
\citet{cristallo09}, \citet{weiss09}, \citet{karakas10b}, \citet{ventura09b},
\citet{ventura10}, \citet{stancliffe10}, and \citet{karakas12}.
The impact of low-temperature opacity tables on the stellar evolution
of LIMS is still being assessed, with many
more studies on this topic expected. Using accurate molecular opacity tables
shortens AGB lifetime, with great impact on PN production.

\subsection{Nucleosynthesis and Abundances}
The same stellar evolution codes are also used to calculate
post-AGB evolutionary sequences \citep{blocker95b}, although stellar 
models calculated self-consistently from the main sequence
through to the white dwarf cooling track are preferred
\citep[e.g.,][]{miller07}. 
The nucleosynthesis patterns of born-again AGB, post-AGB stars, and
PNe are very useful tools for constraining the
evolution and mixing of the progenitors during the AGB and
beyond.  The composition of hot, rare PG 1159 stars has been particularly
illuminating as these are descendants of post-AGB evolution, forming
after a late or very late TP \citep{bloecker01,miller06}. 
The abundances of C and O in particular have been shown to
be much higher than in the composition of standard AGB
models \citep{werner09}. The intershell compositions of PG 1159
stars have been well explained in terms of intershell convection
penetrating into the C-O core \citep{herwig01}, but there is still
debate as to the physics that drives this \citep{stancliffe11}. 
Future work through the use of higher-resolution hydrodynamical models or from
better observational constraints may help settle this. Other stars such as the born-again
Sakurai's Object 
and FG Sagittae 
show real-time stellar evolution 
and challenge our simplistic notion of stellar convection 
\citep{herwig11}. Future hydrodynamical modeling is essential
to an improved understanding of post-AGB evolution.

The abundances of elements heavier than iron are an
exquisite tracer of the thermodynamic conditions in the He-burning 
shells  of AGB stars. Heavy elements are produced by the $s$-process in the
He intershell and mixed to the surface by convective processes.
Unfortunately, accurate determinations of photospheric abundances from
AGB stars are difficult owing to strong molecular opacities and dynamic
atmospheres \citep[e.g.,][]{abia08}. 
While it has become possible to obtain constraints on, e.g., 
the heavy element nucleosynthesis in intermediate-mass AGB stars
\citep{garcia06,garcia09,karakas12}, uncertainties are large. 
It is simpler to obtain abundances from post-AGB stars 
\citep[e.g.,][]{vanwinkel00,desmedt12} but statistics are 
currently rather limited.  Future work will expand the number 
of post-AGB stars with reliable abundances as a result of
large-scale surveys of post-AGB stars in the Magellanic Clouds.

Abundances can be obtained from PN spectra for a number of 
elements such as C, N, O, S, Cl and the noble gases He, Ne, and
Ar (see \S8). The abundances of noble gases and  other elements
(e.g., Cl, Ge, and Br) cannot be derived from the spectra of cool
evolved stars, making the information that comes from PNe
unique. Chemical abundances constrain the initial composition
of the progenitor star and provide clues to mixing and
nucleosynthesis. 

The heavy element composition from PNe is a relatively
new and exciting field, providing data for, e.g., 
Zn, Ge, Se, Kr, Xe, and Ba, some of which are difficult or impossible
to observe in AGB or post-AGB star spectra \citep{pb94, 
dinerstein01a, sterling02,sharpee07,sterling07,sterling08}.
Thus, the chemical 
composition of PNe provide a unique insight into nucleosynthesis
during previous phases \citep{karakas09,karakas10c}. One outstanding
issue is that abundances of heavy elements derived from 
PNe suffer from large uncertainties, driven in part by a
lack of good atomic data. This is currently changing 
\citep[e.g.,][]{sterling11a,sterling11b}, with the hope that more 
reliable and extensive  data will be available in the 
near future (see \S7).

\subsection{Binarity}
It should be emphasized that all of the above deals
with single star evolution. Most stars are in binary systems, where
the orbital parameters  and mass ratio 
determine the evolutionary fate of the system. Binary interactions are
thought to be highly important in shaping PNe
(see \S4) as well as leading to explosive 
phenomena such as novae and Type Ia supernovae. Binary evolution is
horrendously complicated, in particular due to the complexity of the
types of stellar interactions that are possible. For example, if one 
star fills its Roche Lobe, a common envelope can develop. We currently
have a very limited understanding of common envelope evolution,
a situation that is being 
improved through hydrodynamical simulations
\citep{passy12, ricker12}. Future work will need to address how to
implement improvements from hydrodynamical simulations 
into stellar evolution or population synthesis models. 
More work is also desperately needed
to address how binary evolution changes the chemical yields of
single stars. 

\section{The Impact of Binary Interactions on the PN Population}

Quantifying the impact of binarity on the formation and evolution of PNe needs two distinct approaches. With the first, we study individual PN harbouring binary central stars, with the aim of connecting stellar and binary parameters to PN kinematics and morphology. The second approach is to establish to what extent binary interactions have determined the characteristics of the PN population as a whole, by selecting population characteristics (such as mean central star mass) that vary depending on the frequency of interactions.

Currently, we expect that the fraction of PN that have experienced a binary interaction of their central star to be a consequence of the fraction of close binaries in the 1--8M$_\odot$ main sequence population, which is roughly 20-30\% \citep{Raghavan2010}. However, the high incidence of non spherical PNe \citep[$\sim$80\%;][]{Parker2006}, alongside the lack of a quantitative single-star theory that explains these non-spherical shapes \citep{Soker2006,Nordhaus2007}, has prompted questions of whether the PN phenomenon is associated preferentially with binarity. If this were so, by necessity, there would be some AGB stars that never develop a PN because they do not suffer an interaction \citep{Soker2005,DeMarco2010}. 

Here we consider these two sets of problems. For more complete reviews see \citet{DeMarco2009} and \citet{DeMarco2011b}. 

\subsection{Recent  observational efforts to detect binary central stars of PNe and related objects}

To determine how binary interactions have played a role in the formation and shaping of PNe, newly detected binaries \citep[e.g.,][]{Miszalski2009,Boffin2012} need to be characterised, which may include a model of observables such as light and radial velocity curves \citep[e.g.,][]{Hillwig2010}. PN kinematics need then be related to binary parameters \citep[e.g.,][]{Guerrero2012,Mitchell2007,Tocknell2013}. This approach can also be carried out statistically, where a sample of PNe with close central stars is examined for their PN characteristics, as done by \citet{Miszalski2009b}, who concluded that bipolar PNe with filaments tend to be preferentially present around close binary central stars. 

The known binaries are either the very close, post-common envelope ones \citep{Paczynski1976}, detected by periodic photometric variability indicating ellipsoidal distortion, irradiation of eclipses, or are so wide \citep[many thousands to tens of thousands of AU (astronomical units);][]{Ciardullo1999} that the components have not interacted.  We know of $\sim$40 close binaries \citep[e.g.,][]{Bond2000,Miszalski2009} and the close binary fraction is $\sim$15\%. This is expected to be a lower limit because close binaries with unfavourable inclinations, smaller companions, or cooler central stars may not be detected \citep[the sensitive {\it Kepler} satellite has discovered two central star binaries, undetectable from the ground, in a sample of only six;][]{Long2013}(see \S5). 

Central star binaries with intermediate periods must exist.  While at separations of a few solar radii to a few AU there will be no binary central stars because the AGB progenitor would have swept the companion into a common envelope and dramatically reduced the orbital separation, from a few AU to $\sim$100~AU we do expect companions to exist and to have interacted with the AGB star, affecting the mass-loss geometry and possibly the mass-loss rate as well \citep{Soker1997}. Such binaries would have no light variability and only slight radial velocity variability. The first central star binary, with a period of $\sim$3 years was recently detected (van Winckel, in preparation) after a long-term monitoring campaign.

In an attempt to find such elusive companions \citet{DeMarco2013} and Douchin et al. (in preparation) carried out a survey of $\sim$40 central stars in the $I$ and $J$ bands, showing that substantially more than the 20-30\%  expected for the standard scenario may have companions, although only a larger sample can give a more statistically significant answer. 

Finally, based on predictions from the main sequence binary fraction and period distribution we note that the central star {\it close} binary fraction of $\sim$15\% \citep{Bond2000,Miszalski2009} is already much higher than the few percent expected, adopting a maximum tidal capture radius of 2-3 stellar radii \citep{Mustill2012}.

\subsubsection{Binary AGB and post-AGB stars}

Can AGB and post-AGB binaries serve to constrain the binary fraction of PN central stars further? AGB binaries are known \citep[e.g.,][]{Sahai2008,Mayer2013}, but there is no hope for a complete census. In the Magellanic Clouds, half of  the ``sequence E" stars  are thought to be AGB contact binaries \citep{Nicholls2012}, the immediate precursors of post-common envelope PNe. Their numbers are consistent with a lower post-common envelope PN fraction of $\sim$7-9\%, possibly due to the lower age of the LMC population.   

One in three post-AGB stars {\it with no nebulosity} (naked post-AGBs) has been found to have  a companion with a period in the 100-2000-day range and with a Keplerian circumbinary disk \citep{vanWinckel2009}. These objects may not develop a PN, possibly because of re-accretion of material from the disk, which slows the leftward evolution on the Hertzsprung-Russell diagram. These binaries likely went through a strong binary interaction that reduced their period, but not as much as in a classic common envelope interaction. Is it then possible that all strong interactions that avoid the common envelope result in no PN at all? 

Long-term monitoring of a small sample of post-AGB stars {\it with} a pre-PN nebulosity has revealed that at least one, and possibly two, of the monitored objects may be binaries with a decade-long period \citep{Hrivnak2011}. We could therefore conjecture that the binaries in pre-PNe all have decade-long periods. It has been proposed (\citealt{Bright2013}) that it is the wide binary nature of the pre-PN central stars that allows them to have the observed pre-PNe, while the naked post-AGB stars, having suffered a closer binary interaction, experienced re-accretion, which prevents the lighting of the PN. An observational test of this scenario is the characterisation of the disks around naked post-AGB stars as well as those surrounded by a pre-PN, accompanied by a model of their formation \citep{Bright2013}.

\subsection{Recent fundamental theoretical efforts}

Seminal work in PN shaping was carried out by \citet[][see also \citealt{GarciaSegura2005}]{GarciaSegura1999}, who developed a model of how rotation and magnetic fields in AGB stars can lead to mass-loss geometries that later give rise to a range of PN shapes. However, the magnetic fields in those hydrodynamic simulations are artificially constant because gas movement does not feed back to the field strength and direction. Recently, \citet{Soker2006} and \citet{Nordhaus2006} have demonstrated how such feedback would quench the field in approximately a century, eliminating the source of global mass-loss shaping that is needed to obtain non spherical PNe. 

There is broad consensus that the action of a companion can act directly on the AGB envelope gas to generate PN shapes, as well as on the stellar magnetic fields to recreate some of the conditions explored by Garc{\'\i}a-Segura et al. (1999, 2005). However, binary models are few and cover random spots in parameter space with wildly different techniques. \citet{Garcia-Arredondo2004} carried out hydrodynamic simulations of a binary system with a 10~AU separation, the parameter regime discussed analytically by \citet{Soker2000b}, and confirmed that a ``compression" disk is indeed expected when jets are present. However, they did not simulate the conditions that generate the accretion disk that launched the jets in the first place. Recently \citet{hec13} [but see also previous work by \citet{Mastrodemos1998,Mastrodemos1999} and \citet{hef12}], carried out simulations of how the accretion disk forms and cast doubt that the accretion rates needed to blow pre-PN jets in the first place are sufficient if the orbital separation is in the 10-20~AU range. Huarte-Espinosa et al. (2012) compared their work to other papers modeling the formation of accretion disks \citep[e.g.,][]{Mastrodemos1998,deValBorro2009}, but did not even mention the work of Garc{\'\i}a-Arredondo \& Frank (2004), because that paper does not model the accretion disk formation and only models the jets. Yet the conclusions of the two papers are at odds and leave the readers (the observers) to choose which model to follow when contemplating a unified scenario for the observations.

\section{PN Central Stars}

Several important questions concerning PNe are directly related to PN 
central star research. Examples are: distance determinations, binary
fraction, distribution of binary orbital periods, fraction of H-depleted 
photospheres, properties of central star winds. Below we offer a list
of research fields where we can expect significant progress in the near
future.

\subsection{Astrometry}

The epoch of Gaia is coming soon, but it is not completely clear what 
kind of impact Gaia will have concerning trigonometric parallaxes of
central stars. In many cases the relatively low angular resolution
and observing mode may make it 
difficult to work with images of stars embedded in bright nebulae.
(It might be useful to request, from some authoritative Gaia source,
like CU4, information on what they are planning to do about PNe. 
{\it Ad hoc} simulations demonstrating performance, if not planned yet, 
could be suggested. 
Eventually our Working Group might offer a list of targets we consider 
particularly important for PN research; see below on ``spectroscopy").

In the meantime, other trigonometric parallax measurements are 
steadily improving. The latest PN results are from {\it HST} Fine Guidance 
Sensor data \citep{b09}, offering distances as large as 
600 pc. But especially ground-based techniques using IR adaptive optics, 
as in \citet{lgy10} or \citet{cbk09}, appear to be able to
produce, at least in favorable stellar fields, trigonometric
parallaxes below 1 milli-arc second, which means distances larger 
than 1 kpc. Central star parallax determinations using infrared adaptive optics should be 
encouraged, because even a few measurements would be astrophysically
important. 

\subsection{Photometry}

Photometric studies of central stars have two important applications: the
search for (a) binary systems, and (b) surface instabilities produced by
pulsations and stellar wind variations. In the near future, new sources
of photometric information will become available. The {\it Kepler} field has
already produced some evidence that a large fraction of PN central stars
may show small amplitude photometric variations \citep{djd12}.
All-sky multi-filter surveys like Pan-STARRS\footnote{Panoramic Survey Telescope and Rapid Response System} are being carefully
calibrated to provide photometry with accuracy at or below the 1\% level
\citep{tsl12}.
Systematic searches for PN central star photometric variations in these 
and other similar databases are being planned, and can be expected to 
help us understand the photometric behavior of central stars in much 
more detail than has been possible up to now.

\subsection{Spectroscopy}

There are at least three excellent reasons to want central star spectra:
 
(1) They give information about basic stellar photospheric and wind 
parameters, allowing us to test radiatively-driven wind models and 
stellar structure and evolution models;

(2) In particular, they provide photospheric abundances, which is
a good way of testing post-AGB model predictions, starting with the 
fraction of H-deficient central stars; and

(3) They provide radial velocities, which will eventually provide an irreplaceable 
way of refining our knowledge of the binary orbital period distribution
and binary fraction among PN central stars.

It should be clear, however, that what we need are spectrograms
with high spectral resolution and the highest possible signal-to-noise 
ratio. Even spectral classification should be based only on high-resolution 
spectra, because, at low resolution, the presence of nebular Balmer 
emission lines does not allow us to judge whether or not the central star
has an H-deficient photosphere (which is arguably the most important
information we should expect a spectral type to give). 

This point requires more explanation. Since low-resolution spectra are 
easier to obtain, many authors tried to base their spectral descriptions
on such insufficient data. This resulted in the introduction of the {\it``wels"}
(weak emission line star) description, which ignored the basic question
of H-deficiency. There has been a tendency to interpret the {\it wels} label
as a spectral type. We have even seen stars, previously classified as
H-rich from high-resolution spectra, ``reclassified" as {\it wels} from
low-resolution spectra (authors and referees were equally guilty).
For this reason it is necessary to emphasize that 
{\it wels} is not a spectral type, and and that its use should be restricted 
to mean ``insufficient spectral resolution.'' The point has been made 
again by \citet{mi12} in the context of close binary searches.

So the challenge is how to obtain the large amount of telescope time 
required for high-resolution spectrograms; and even more daunting, 
how to take many such spectrograms, 
in a search for radial velocity variations. Probably the best strategy 
is to try to join forces with other groups requiring large 
amounts of high-resolution spectra: in particular, exoplanet searchers.
Central star spectroscopy should not be restricted to the visible range; 
efforts to obtain UV spectra, for example with the {\it HST} Cosmic Origins 
Spectrograph, would be very welcome.

Traditionally, only H and He photospheric central star abundances
could be determined, because of complications related to the high 
surface temperature (requiring a non-LTE treatment of the atmosphere 
as a whole, and in particular of metal line opacity), and the presence 
of significant mass loss (requiring hydrodynamically consistent winds
in an extended, expanding atmosphere). Modern model atmospheres are
reaching a level of development that allows us to extend photospheric 
abundance determinations to several other elements, like C, N, O, Si, 
P and S; see, e.g., \citet{mg09}. However, 
there are some difficulties. The number of physical parameters 
to be fitted is now larger (surface temperature, log $g$, abundances, 
mass loss rate, wind terminal velocity, velocity law, clumping filling 
factor). It is particularly in this respect that UV spectrograms provide 
essential complementary information; see, e.g. \citet{hb11}.
 
There is also the potential capability of deriving stellar 
radii, masses and luminosities from wind properties, without depending 
on the luminosity -- core mass relation. In practice, the current state 
of wind models leads to an unresolved conflict: the masses of central 
stars derived from hydrodynamically consistent wind theory disagree 
with the masses we would derive from adopting the luminosity -- core 
mass relation, as reported by \citet{kph12}. 
This kind of problem can only be solved by measuring reliable
distances, like those from trigonometric parallaxes. The known distance 
then fixes the stellar radius and mass unambiguously. A few PN central 
star trigonometric parallaxes (in particular for those stars showing a 
strong wind) could have a strong impact on stellar wind theory and 
stellar evolution theory.

If we can grow confident in the accuracy of photospheric abundances
derived from the newest generation of model atmospheres, then a new
way of testing and forcing predictions from stellar evolutionary 
theory will become available.

\section{Planetary Nebula Morphology and Kinematics}

\subsection{Morphology}

The study of the PN morphology has played a
fundamental role in their understanding.  PNe display a myriad of
shapes that early observers grouped into fundamental classes. These
basic morphological classes have later been related to physical
properties such as the mass of the progenitor star, the chemical
yields in the expelled stellar material and the galactic distribution
of the population of PNe (e.g. Torres-Peimbert \& Peimbert 1997).

The advent of high spatial resolution imaging, initiated with the 
WFPC~2\footnote{Wide Field Planetary Camera~2} camera on board {\it HST}, followed
by adaptive optics systems in large ground-based telescopes, have
revealed additional complex structures and outstanding symmetries
since the first stages of development of the PN phase 
(e.g. Sahai, Morris \& Villar 2011).  The structure and substructures of a 
PN are related to its remnant core mass, environment and nuclear conditions 
at the exit of the AGB and the subsequent mass-loss and photoionization 
evolution.

The presence of thick toroidal structures or equatorial mass
enhancements and the presence of symmetric structures, either compact
as knots or elongated as collimated lobes, have led to the inference that the
core may be often composed by a close binary system (e.g., De Marco 2009; 
L\'opez et al. 2011 and references therein).

\subsection{Kinematics}

Since the morphology evolution of a PN depends on the outflowing
properties of the gaseous shell, the study of kinematics and
morphology in PNe have developed closely in parallel 
(e.g. L\'opez et al. 2012a; Garc{\'\i}a-D{\'\i}az et al. 2012; Pereyra et al 2013;
Huarte-Espinoza et al. 2012).
 In addition, the mean or systemic radial velocity of each PN is a 
key ingredient in
understanding the population distribution of PNe in the galactic
context, locally and in other galaxies (e.g. Durand, Acker \& Zijlstra 1998). 
 For the local population,
spatially resolved kinematic information of the different structures
that compose a PN (shell, rim, lobes, halo, etc) provide a detailed
insight into the global shell dynamics (L\'opez et al. 2012b). 
High-speed, collimated,
bipolar outflows are now known to be a relatively common occurrence in
PNe and their origin has also been considered as likely originating in a
binary core with magnetic fields providing collimation conditions 
(e.g Garc{\'\i}a-Segura and L\'opez 2000).

Moreover, since the measured velocity at a certain point in the nebula
corresponds to the magnitude of the radial component of the velocity
vector, this information provides a good insight into the structure of
the nebula along the line of sight, considering that the expansion is
homologous or does not depart far from it. In this way a reasonably
good 3D reconstruction of the nebular shape can be achieved for objects
whose 2D projection on the sky (the image) is known and where good
kinematic mapping over the face of the nebula is available (Steffen 
\& L\'opez 2006).

\subsection{Future prospects}

Current instrumentation provides access to nearly all relevant
areas of the electromagnetic spectrum. Imaging at different
wavelengths at high spatial resolution will provide a much better
understanding of the interplay among the different contributors to the
PN phenomenon throughout its evolution, from the hot plasma in the
X-ray domain, the photoionized gas in the optical and radio regimes to
the warm and cool dust and molecular component in the infrared and
millimeter ranges.

In the optical and the infrared regimes, morphology and kinematics can
now be obtained simultaneously through the use of integral field
spectrographs (IFS) with nearly equivalent mapping quality in both
cases (only depending on the spaxel aspect ratio) and the data,
morphology and kinematics, analyzed comprehensively point to point as
a function of wavelength from the data cubes. Since many of the
current IFS are operating on large telescopes, many of these have been
designed to work in the near infrared, with more expected to become
operational in the near future (e.g., L\'opez et al 2007). This means 
that the warm and
photo-dissociated regions in very young PNe will become better
studied. This is a highly relevant stage to understand the early
formation and development of the PN shell and one that has not been
studied in the depth it deserves. IFS data lead in a natural way to a
3D reconstruction of the nebular structure and location of the stellar
component(s) embedded within it. These data will surely be fruitfully
complemented by the data cubes from more millimeter observations in
the near future, particularly from {\it ALMA} (Atacama Large Millimeter/submillimeter Array) (e.g., Bujarrabal et al. 2013). 
Combining high spatial and
spectral resolution data cubes at multiple wavelengths and deriving
from them the 3D core structure and outflows that produce PNe will be
the ultimate challenge in the coming years. 

This information will be often complemented by the measurement of the
motion of the gas or dust in the plane of the sky, thanks to the
increasing time baseline for which high-resolution images of
individual nebulae are becoming available (e.g., Balick \& Hajian 2004;
Corradi et al. 2011b; Balick et al. 2013). The combination of
multiple-epoch images and Doppler shift velocities will allow to
gather information on five out of the six dimensions of the
phase-space, and in some cases even detect acceleration (hence the
forces at work), providing unique new constraints to the
hydrodynamical modeling.

For quasi-spherical nebulae, it seems that a reasonable understanding
of the dynamical evolution of the different observed shells has been
reached (Perinotto et al. 2004 and the following articles of the
series). In this respect, we stress the potential of
haloes to reconstruct the mass loss history of the progenitors stars
in the last 10$^5$~yr of the AGB (Corradi et al. 2003). The rings/arcs
often observed in the inner regions of the haloes (Corradi et
al. 2004) are instead poorly understood but potentially important, and
call for further investigation.

For collimated outflows, the main goal will be to recover accurately
both the mass loss history and geometry of the progenitors, providing
conclusive evidence of the physical phenomena suggested so far, such
as fast/collimated/pre\-ce\-ssing/intermittent winds or bullet-like
ejections (e.g., Balick et al. 2013).

The dynamical properties should also be more tightly related to the
physics of the central sources. As mentioned before, binarity is
supposed to be a key ingredient to explain many of the observed
morphologies. While the number of binary central stars is rapidly
increasing (Miszalski et al. 2009a), a clear correlation with
morphology has been demonstrated only for few, specific components
such as extreme equatorial rings and polar outflows (Miszalski et
al. 2009b; Corradi et al. 2011a; Boffin et al. 2012). A more
comprehensive understanding of the most common basic geometries of the
nebulae is still lacking.

The role of magnetic fields, often invoked in the modeling of the
nebulae, should also be better clarified. So far no evidence exists
for magnetic fields in the central stars (Asensio-Ramos et al. 2013,
in preparation), and little information is available for the nebular
gas.  Polarimetric studies from the IR to radio wavelengths, revealing
dust polarization (Sabin et al. 2007), and in the optical/UV searching
for fields anchored to the central stars, are needed for further progress.

\section{Atomic processes}

\subsection{Introduction}

PNe are ideal `astrophysical laboratories' for studying
atomic processes. The basic atomic processes in PNe are photoionization and
recombination (including radiative and dielectronic recombination),
charge-exchange reaction, collisional excitation and de-excitation, and
radiative transition. Today atomic data for these processes are all available
and, for some ions, have been calculated to high accuracy. Much effort is
still needed to revise the old data as well as to create new data.

\subsection{Atomic data for collisionally excited lines}

The optical and far-infrared fine-structure transitions within the ground
configurations of O$^{2+}$ and N$^{+}$ are standard tools for nebular analysis
(Liu et al. \citeyear{liu01a}; Osterbrock \& Ferland \citeyear{of06}).
Although O$^{2+}$ has been studied extensively, recent calculations show
that taking relativistic effects into account will change the collision
strengths by 20\% (Palay et al. \citeyear{pal12}). Thus the calculation
of O$^{2+}$ needs careful revision. Accuracies of the N$^{+}$ collision data
are claimed to be 20\% (Tayal \citeyear{tay11}), however, using
different target wave functions results in significant differences in the
collision strengths. Efforts are needed to improve the atomic model.

Studies of the [O~{\sc ii}] $\lambda$3729/$\lambda$3726 ratio at the
high-electron-density limit suggest that large-scale electron correlations
and quantum electrodynamics corrections should be considered (Han et al.
\citeyear{han12}). Future theoretical work on O$^{+}$ needs to use increased
atomic orbitals and take into full account relativistic effects. The radiative
transition rates and collision strengths of S$^{+}$ given by Tayal \&
Zatsarinny \citeyearpar{tz10} are claimed to be better than 30\%. Given
the complexity of the e-S$^{+}$ scattering, further investigation is desired.

Recent theoretical work on Ne$^{2+}$ by Landi \& Bhatia \citeyearpar{lb05}
shows substantial differences in transition probabilities compared with
previous calculations and experiment, due to a larger number of configurations
used in the new calculation. Further investigation of the Ne$^{2+}$ atomic
model is needed before a conclusion can be drawn. New collision strengths for
the fine-structure transitions of S$^{2+}$ (Hudson et al. \citeyear{hud12})
differ significantly from an earlier calculation. More extensive study of
e-S$^{2+}$ scattering, with emphasis on the near-threshold resonances, is
needed.

The transition probabilities and collision strengths of Ne$^{3+}$ currently
used are from Zeippen \citeyearpar{zei82} and Ramsbottom et al.
\citeyearpar{ram98}, and are out of date. For Ar$^{3+}$, those data are from
Mendoza \& Zeippen \citeyearpar{mz82} and Ramsbottom et al.
\citeyearpar{ram97}. More sophisticated calculations for the two ions are
needed. The transition probabilities of Cl$^{2+}$ given by Sossah \& Tayal
\citeyearpar{st12} agree well with experiment and earlier calculations, but
collision strengths show significant discrepancies. Future studies of
Cl$^{2+}$ need to consider the effects of coupling to the continuum, which
may be important for some transitions, but is ignored in the current work.

\subsection{The iron-peak elements}

Emission lines of Fe$^{2+}$ and Fe$^{3+}$ have been extensively studied
(Rubin et al. \citeyear{rub97}; Rodr\'{i}guez \citeyear{rod03}; Rodr\'{i}guez \&
Rubin \citeyear{rr05}). Nahar \citeyearpar{nah06} and McLaughlin et al.
\citeyearpar{mcl06}, respectively, presented new radiative transition
probabilities and collision strengths of Fe$^{3+}$, and varying degrees of
agreement were found in transition probabilities compared with earlier
calculations. The transition and collision data of Fe$^{2+}$ currently in use
are from Nahar \& Pradhan \citeyearpar{np96} and Zhang \citeyearpar{zha96}.
New calculations of Fe$^{2+}$ are needed. Atomic data for other elements in
the iron-peak group are still limited, with most work mainly focused on
singly-ionized ions (Bautista \citeyear{bau06}). Future investigation should
extend to doubly and triply ionized ions of Sc to Co.

\subsection{Neutron-capture elements}

Since the seminal work of P\'{e}quignot \& Baluteau \citeyearpar{pb94},
emission lines of neutron($n$)-capture elements have been identified in
nearly 100 Galactic PNe. Accurate abundance analysis using those lines relies
on accurate atomic data. The calculations of Sterling \& Witthoeft
\citeyearpar{sterling11a}, Sterling \citeyearpar{sterling11b} and Sterling \& Stancil
\citeyearpar{ss11} enable determination of the Se and Kr abundances and
investigation of the $s$-process enrichment in PNe. However, nebular
astrophysics still lacks of atomic data for the $n$-capture elements. The
current atomic models for heavy elements need to be revised and fully
relativistic atomic code should be developed. The  sensitivity of abundance determination
to uncertainties in atomic data of the trans-iron elements also needs to be
tested (Sterling et al. \citeyear{ste12}).

\subsection{Hydrogen}

The recombination spectrum of hydrogen has been extensively used in studies
of nebulae (Peimbert \citeyear{pei71}; Barker \citeyear{bar78}; Liu \& Danziger
\citeyear{ld93}; Liu et al. \citeyear{liu00}; Zhang et al. \citeyear{zha04}).
The most accurate calculations of H~{\sc i} recombination lines is by Storey
\& Hummer \citeyearpar{sh95}. Unresolved problems in hydrogen are:
1) Radiative transfer of Ly$\alpha$ photons including the effects of
self-absorption, removal and collision-induced transitions between the
2s\,$^2$S and 2p\,$^2$P$^{\rm o}$ states (Bautista \citeyear{bau06});
and 2) collisional excitation from the ground state to $n\,>\,2$ states under
high-$T_\mathrm{e}$ and high-$N_\mathrm{e}$ conditions, which affect 
primordial helium abundance determinations (Peimbert et al. \citeyear{pei07}).

\subsection{Helium}

Theoretical He~{\sc i} line emissivities are now as good as 1\%
(Bauman et al. \citeyear{bau05}; Benjamin et al.
\citeyear{bss99},\,\citeyear{bss02}).
Plasma diagnostics based on the He~{\sc i} lines are established (Zhang et
al. \citeyear{zha05a}). Future work will concentrate on 1) collision
cross-sections of He~{\sc i}; 2) optical depth effects on the He~{\sc i}
$\lambda$3889 line; 3) self-absorption due to significant population of the
1s2s\,$^3$S$_{1}$ meta-stable state; and 4) mechanisms that may destroy the
He~{\sc i} resonance line photons, such as photoionization of neutral hydrogen
or absorption by dust grains (Liu et al. \citeyear{liu01b}).

\subsection{Effective recombination coefficients for heavy element optical recombination lines}

\subsubsection{The second-row elements}

Heavy element optical recombination lines (ORLs) have proved to originate from very `cold' regions
(e.g., Liu et al. \citeyear{lbzbs}; McNabb et al. \citeyear{mcn13}; Fang \& Liu
\citeyear{fl13}), in line with the bi-abundance nebular model (Liu et al.
\citeyear{liu00}). In order to study the properties of such `cold'
inclusions, reliable effective recombination coefficients for the C~{\sc ii},
N~{\sc ii}, O~{\sc ii} and Ne~{\sc ii} lines are needed.

The commonly used C~{\sc ii} effective recombination coefficients of Davey et
al. \citeyearpar{dsk00} were calculated in {\it LS}~coupling at a single
electron density of 10$^4$~cm$^{-3}$. More extensive calculations for this
ion should be carried out in intermediate coupling and cover a broad range
of temperatures and densities. Sochi \citeyearpar{soc12} presents new
intermediate coupling calculations of C~{\sc ii}, but only includes the
dielectronic recombination.

New calculations of the effective recombination coefficients for the N~{\sc 
ii} lines (Fang et al. \citeyear{fsl11},\,\citeyear{fsl13}) and O~{\sc ii} lines (Storey, in preparation) are a great improvement over previous work
(Storey \citeyear{sto94}; Liu et al. \citeyear{lsbc}; Kisielius \& Storey
\citeyear{ks02}). Plasma diagnostics based on the N~{\sc ii} and O~{\sc ii}
ORLs are constructed (McNabb et al. \citeyear{mcn13}). However, problems still
exist in calculations, e.g., compared to earlier work, the N$^{2+}$ target
state energies generated by Fang et al. \citeyearpar{fsl11} agree less well
with experiment; level population calculations of N~{\sc ii} become more
uncertain at very low-$T_\mathrm{e}$, high-$N_\mathrm{e}$ conditions. Future
improvement includes: 1) creation of an N$^{2+}$ target with larger
configuration set; 2) a full collisional-radiative treatment among the low
states; and 3) ab initio calculations extended to higher $n$.

The inadequacy of the currently used Ne~{\sc ii} effective recombination
coefficients calculated by Kisielius et al. \citeyearpar{kis98} was noticed
through very deep spectroscopy of the Galactic PN NGC\,7009 (Fang \& Liu
\citeyear{fl13}). A new calculation for Ne~{\sc ii} needs to 1) consider that
relative populations of the Ne$^{2+}$ $^3$P$_{2,1,0}$ fine-structure levels
deviate from the statistical weight ratio $5:3:1$; 2) be carried out in
intermediate coupling; and 3) be extended to temperatures well below 1000~K,
including the effects of fine-structure dielectronic recombination.

\subsubsection{The third-row elements}

The only atomic data available for the Mg and Si ORLs are the dielectronic
recombination coefficients given by Nussbaumer \& Storey \citeyearpar{ns86},
whose calculations were restricted to those ions for which the recombining
ions have $n$=3 valence electrons. Thus Mg~{\sc ii} and Si~{\sc iv} as well
as ions with higher ionization stages were not calculated. Among all the
third-row elements, the effective recombination coefficients for the Mg~{\sc
ii} and Si~{\sc iv} lines are the most needed (Barlow et al. \citeyear{bar03}; Liu
et al. \citeyear{liu04}; Tsamis et al. \citeyear{tsa04}; Zhang et al.
\citeyear{zha05b}; Liu \citeyear{liu06}; Fang \& Liu \citeyear{fl13}).

\section{Abundance Determinations}
Abundances in PNe reflect both the nucleosynthesis contributions of their progenitor
 stars and the chemical signatures of their galactic birthplaces. Thus the abundance profiles of PNe, individually
  and collectively, contain valuable information about galactic chemical evolution. Over the last two decades the
   study of the chemical compositions of PNe has been extended beyond the abundant elements in the first two
    rows of the periodic table and the exclusive use of strong forbidden lines. More recently, the development
     of new observational capabilities and analytical techniques has highlighted the challenges and potential
      rewards of accurate, reliable abundance determinations. Listed below are some prospects for further advancement.

{\subsection{The Forbidden-Permitted Discrepancy}
Among the most persistent problems dogging PN abundance determinations is the disagreement between 
abundances calculated from collisionally-excited (forbidden) lines and those from permitted recombination 
lines. Re\-com\-bi\-na\-tion-line abundances are systematically higher than those derived from forbidden lines, 
often by a large factor. Models based on fluctuations in local conditions within chemically homogeneous 
nebulae vie with those incorporating a bimodal abundance distribution to explain these discrepancies 
(see e.g., \citealt{P11}). Observational progress in detailing the temperature, 
density, and abundance structures in PNe requires spatially-resolved spectra that 1) are deep enough to 
unambiguously detect faint recombination lines; and 2) possess sufficient resolution to minimize the 
confounding effects of line blends. New capabilities in this area are represented by MEGARA\footnote{Multi-Espectr{\'o}grafo en GTC de Alta Resoluci{\'o}n para Astronom{\'i}a}, an 
IFU/MOS\footnote{integral field unit/multi-object spectrograph} instrument scheduled for first light on the Gran Telescopio Canarias in 2015 \citep{gdp12}, 
and MUSE\footnote{Multi Unit Spectroscopic Explorer}, an adaptive-optics IFU planned for the VLT late in 2013 (as of this writing the instrument is undergoing reintegration in Paranal).

Spatially-resolved maps of infrared 
 fine-structure lines as observed with {\it SOFIA}\footnote{Stratospheric Observatory for Infrared Astronomy} and {\it Herschel} could supply further evidence of the existence of the proposed cold, H-deficient inclusions, if these lines are found to peak in the regions exhibiting the 
 highest abundance discrepancy \citep{rub12}.  

On the theoretical side, \citet{nds12} suggest that if free electron energies deviate 
from a  Maxwell-Boltzmann distribution, and instead follow a {\it kappa} distribution, this may explain the 
abundance discrepancy. Studies such as that by \citet{ss13}, who investigated 
the free-electron distribution from C~II lines in a sample of PNe, will help to determine if this is a viable interpretation. 
If so, then the challenge turns to incorporating {\it kappa} distributions into existing modeling codes and understanding 
how these distributions behave as the nebula evolves.  

\subsection{Neutron-capture elements} 
Expanding the inventory of elements detected in PN gas augments our ability to make useful comparisons between 
observed abundance patterns and those predicted by theory. Of the less-abundant elements, s-process nuclei 
(e.g., Se, Br, Kr, Rb, Sr, Te, I, and Xe) are especially useful, as their relative abundances can provide robust 
constraints on computational models of AGB evolution (particularly during the late stages of the thermally-pulsing 
phase). Measuring s-process abundances requires echelle-level resolution of faint lines along with accurate 
methods for untangling blends, so studies of these elements will benefit from the availability of instruments 
mentioned above. It is also possible, using large ground-based telescopes, to detect neutron-capture elements 
in the PNe of neighboring galaxies. Results of such investigations can illuminate the dependence of s-process 
enrichment factors on both metallicity and progenitor mass.
 
\subsection{Ionic to Total Abundances}
Without observations of lines from all populated ionization states, the derivation of total element abundances from 
observed ionic abundances relies on either full photoionization modeling (often with incomplete information), 
or the application of ionization correction factors (ICFs), or a combination of the two. Published sets of ICFs 
are in general use (e.g., \citealt{kb94,kh01}), 
but the universality of a single ICF for a given element is not necessarily supported; improvements in understanding 
the dependencies of ICFs may point toward the need for customized formulations.  For example, \citet{psbt12}
 find that ICF-derived neon abundances vary with both metallicity and ionization level. 
Even for oxygen, the ``gold standard" element in photoionized nebula abundance determinations, a universal 
ICF does not produce the most accurate results. \citet{mhf12} 
used a grid of photoionization models to study the oxygen ICF over a broad range in central star temperature, 
yielding different ICFs for three stellar temperature regimes spanning 31,000 -- 500,000~K. Finally, sulfur has 
long been troublesome (e.g., \citealt{hsk12}), with ICF-derived abundances for PNe falling systematically 
below those of H~II regions at the same metallicity, with increasing deficits at low metallicity. The difficulty in 
devising a suitable ICF may arise from hydrodynamical effects enhancing higher-ionization states, especially 
at lower metallicities, as suggested by the 1D-RHD (radiation hydrodynamics) results by \citet{jsss12}. Such RHD 
models represent a powerful tool in confronting nebular observations in general, and in assessing ICF reliability, 
in particular. ICFs can also be affected by PN morphology, as suggested by 3D modeling results for bipolars and 
ellipticals \citep{gwm12}; additional studies of sufficiently high-surface-brightness 
objects will help to constrain ICF behavior.  
  
\subsection{Future Directions}
 Future abundance determinations will benefit from improvements in atomic physics calculations at a wide 
 range of physical conditions (including transition probabilities, effective collision strengths, photoionization 
 cross sections, and rate coefficients for radiative recombination, dielectronic recombination, and charge 
 transfer (see \S7), ideally benchmarked against experimental determinations and, importantly, including estimates of uncertainty. 
 
 Research groups involved in PN abundance determinations should be in communication, especially as new atomic data are published. Meaningful comparison of abundances derived by different groups is not always straightforward, and must include ramifications of 
using particular atomic data and varying techniques, as well as understanding the sources of uncertainty. A recent tool addressing this last issue is the NEAT code of \citet{wss12}, which uses a Monte Carlo propagation of line flux uncertainties to estimate uncertainties in derived abundances.  

\section{The properties of dust in Planetary Nebulae}

Low- and intermediate-mass ( M$<$8 M$_{\rm \odot}$), stars
constitute the majority of the stars in the Universe. Many of these
stars end their lives with a phase of strong mass loss and
experience thermal pulses on the AGB,
representing one of the main contributors to interstellar medium
enrichment. As the star evolves through the last phases of the AGB,
mass loss increases up to 10$^{-4}$ M$_{\rm \odot}$
\citep{Herwig05}, forming a circumstellar envelope around the star.
As this envelope expands, it eventually cools down and large number
of molecules are formed \citep{Olofsson97} along with solid-state
species \citep{kwok04}. Depending on the mass of the progenitor
star, this envelope will be either carbon or oxygen rich. For
Galactic PNe, they will be C-rich for M $<$ 4 M$_{\rm \odot}$ and
O-rich for M $>$ 4 M$_{\rm \odot}$ \citep{MDV99}, and for the
Magellanic cloud, C-rich for M $<$ 3 M$_{\rm \odot}$ and O-rich for
M $>$ 3 M$_{\rm \odot}$ \citep{VDM2000}. 

\subsection{Molecular Composition}
Recent advances
in infrared instrumentation, e.g. {\it ISO, Spitzer, Akari, Herschel},
have allowed some of the resulting species to be identified. Among the
inorganic species, the most prominent are the amorphous silicates,
with absorption features at 9.7 and 18 $\mu$m, and crystaline
silicates, such as olivines and pyroxenes, which have
absorption/emission features from 10 to 45 $\mu$m
(\citealt{Sylvester99,GH07}). In addition, the broad 10-15 $\mu$m 
emission (centered at about 11.5 $\mu$m) and the 25-35 $\mu$m 
emission (the so-called 30 $\mu$m feature) are usually
attributed to SiC, \citep[e.g.,][]{Speck09} and MgS, \citep[e.g.,][]
{Hony02}. However, these observed broad features are quite
consistent with the variety of properties of hydrogenated amorphous
carbon (HACs). From laboratory work, \citet{Grishko01} showed that
HACs can explain the 21, 26, and 30 $\mu$m features, while
\citet{Stan12}, analyzing a sample of compact PNe, classified these
features as aliphatic dust. Therefore, this is an open question that
should be addressed in the future. 

As for organic compounds,
there are a rich variety of aliphatic, aromatic and more complex
molecules such as graphene, fullerenes and bucky-onions. Unidentified
infrared bands (UIB) at 3.3, 6.2, 7.7, 8.6, and 11.3 $\mu$m were
attributed to PAHs, \citep{ATB85}.
However, recently \citet{KZ11} challenged this finding, showing that
the unidentified infrared emission (UIE) bands are consistent with amorphous organic solids with a
mixed aromatic-aliphatic structure. They argue that this mixture is
similar to organic materials found in meteorites. More recently
\citet{CGM13} presented new laboratory data showing that the UIE
bands in the C-H stretching region (~3-4 $\mu$m), can be reproduced
by the petroleum fraction BQ-1, which exhibits a mixed
aromatic/aliphatic and cycloaliphatic character. The far-IR bands
are best reproduced by a petroleum fractions made by a `core' of two
or three condensed aromatic rings which are extensively alkylated by
aliphatic and cycloaliphatic fractions.

A 21 $\mu$m feature was first discovered by \citep{KVH89}. Several
attempts to identify the carrier of this feature, such as TiC
clusters, have failed  \citep[e.g.,][]{Chigai03}. \citet{CGM13} showed
that a mixture of petroleum fraction BQ-1 with acenes PAHs (e.g.
tetrahydronaphthalene, anthracene, etc.) could contribute to this 21
$\mu$m band.

Large organic molecules, graphene \citep{GHC2011}, and the
fullerenes C60 and C70 \citep{Cami2010, GHC2010} were
recently found in PNe. \citet{GH2012}  and \citet{M2012} proposed
that these fullerenes are formed by UV photochemical processing of
HACs.

Recently \citet{GHD2013} found the largest organic molecules
(bucky-onion, C60@C240 and C60@C240@C540) in the PN TC1.

Additionally, water-ice features at 3.1, 43, and 62 $\mu$m have
been observed in heavily obscured post-AGB stars
(\citealt{Dijkstra2006,Manteiga2011}).

As the envelope of the AGB star expands, it contributes to the
enrichments of the ISM, and eventually contributes to star-forming regions. 
\citet{Trigo09} has shown from the isotopic composition of 
$^{26}$Al, $^{41}$Ca, $^{60}$Fe, and $^{107}$Pd in some of 
the remnants of the primeval
solar system cloud (Calcium-Aluminum-rich inclusions (CAIs) in
chondrites) that our solar system was contaminated by a massive AGB.
Indeed, \citet{PA2002}, found the 3.4 $\mu$m aliphatic features in the
Murchison meteorite. This suggests that the dust composition of our
primeval solar system cloud may share similarities with the dust
around PNe.

\subsection{Future perspectives}

Future research should be carried out in order to have a more
complete scenario for the formation of the different species in the
dust around AGB stars and PNe. In particular, how the different
silicates (both amorphous and crystalline) are formed needs to be
understood. 

As for organic chemistry, a rich variety of molecules
need to be studied. The mechanism involved in the formation of
aliphatic, aromatic and more complex molecules should also be
understood. 

In the next decade, new laboratory work will shed fresh
light on these questions and help lead to their resolution.
The {\it Herschel} database will expand these studies to the far-IR
spectral range.
{\it ALMA} will open a unique window that will allow us to search for new species.
Space probes will bring new information to our understanding of the chemical composition 
of our primeval solar system cloud, that could be extrapolated to the dust around PNe.

\section{Extragalactic Planetary Nebulae}

\subsection{Chemical abundances}

Extragalactic PNe have been searched for in galaxies of all Hubble types because 
their study can shed light on a series of astrophysical problems. In recent years, much work has been devoted to determining chemical abundances in these nebulae as this knowledge helps us to understand, for example, galactic chemical enrichment with time, metallicity-galactocentric distance gradients and their temporal evolution, nucleosynthesis and evolution of central stars in different environmental conditions and at different metallicities. 
In addition, PNe in spiral and irregular galaxies give us the opportunity to analyze the chemical behavior of the ISM at the time of PN formation (from a few Gyr  up to several Gyr ago), by determining  abundances of elements not affected by stellar nucleosynthesis. The comparison of these abundances with that of the present ISM (represented by H~II regions, HIIs), allows us to study the chemical evolution of the ISM, and  to constraint chemical evolution models of galaxies.

A recent review describing several results on these subjects can be found in \citet{msg12}.

\subsubsection{The available data for different galaxies and their analysis}

Spectrophotometric data for extragalactic PNe, in many aspects, allows a more confident analysis than for Galactic PNe since, in distant galaxies, usually all the nebula is included in the slit, and PN distances to the galactic center are better known than they are in the Milky Way. But PNe in external galaxies are faint objects and their chemical abundances are difficult to determine with precision, as they are based on the detection of the very faint lines necessary for plasma diagnostics. Thus, although large numbers of PNe have been detected in many galaxies, abundances have been determined in only a handful of them, mostly  in Local Group galaxies and only a few farther away.
 
Therefore, we need to substantially improve the amount and quality of data. New surveys of PNe and deep follow-up spectroscopy (especially of the faint ones in order to define the early and late PN evolution) with large-aperture telescopes are crucial.

 \subsubsection{PNe in dwarf galaxies}

In the Local Group and its neighborhood,  PN abundances have been determined for several irregular galaxies: the Magellanic Clouds,
IC10, NGC6822, Sextans A and B and NGC3109 (see \citealt{ld06,msg12,hcpp11,mlc05,kgp05,psr07}). PNe in the dwarf spheroidals NGC185, NGC205, M32 and NGC147, companions of M31, Fornax and Sagittarius galaxies have also been studied (see \citealt{gmm12,rm95,rsm99,kzg08} and references therein).

PN analysis in such galaxies has been used to study the evidence and efficiency of the third dredge-up (TDU) which carries O and other elements to the stellar surface. O and Ne abundances in PNe, compared to HIIs abundances, show that TDU has occurred in many PNe at low metallicities (log O/H $\leq$ 7.8. PNe in NGC\,3109 and the one in Sextans A  are the best examples). This is predicted by some stellar evolution models (e.g., \citealt{k10,m01}), however not  all metal-poor PNe  show evidence of TDU. Thus, the efficiency of TDU seems to depend not only on metallicity but other phenomena are apparently playing a role too. Chemical analysis of a larger sample of PNe at low metallicities should improve our knowledge of TDU and provide better constraints for stellar evolution models (which should include more complete physics) and yield calculations.

Similarly, the chemical behavior of PNe and HIIs in dwarf galaxies have also been used to test chemical evolution models of galaxies, such as IC\,10 (Yin et al. 2010) and NGC\,6822 \citep{hcpp11}. Although a ``best fit'' model is obtained in each case, the results are still not satisfactory. For IC\,10, Yin et al.'s best models required a selective galactic wind stripping mainly the heavy elements, which is an ad-hoc solution difficult to justify. For NGC\,6822, \citet{hcpp11} computed models constrained by data from two different-age PN populations and from  HIIs, and they found that there is not a unique model fitting the data. Different combination of parameters (standard or selective winds, different upper mass limit of the IMF, and different yields) can reproduce  the observations adequately. So, more and better observational constrains are needed to properly test the evolution models. Robust abundance ratios such as N/O, Ne/O and He/H in bright and faint PNe are crucial in determining the stellar populations giving rise to PNe.  In addition, the yields provided by stellar evolution models need to be revised: for instance, no model in the literature predicts N yields in agreement with the observations.

\subsubsection{Chemical gradients in spiral galaxies}

Radial abundance gradients derived for PNe, compared to HIIs and stellar gradients for several elements (O, N, Ne, Ar and S), have been analyzed in
some spirals: M31, M33, M81, and NGC 300 (see \citealt{klb12,bsv10,msv09,smv10,spb13}). (Certainly, gradients in the Milky Way have been studied by many authors, [e.g., \citet{hkj10,sh10}], although in this case the uncertainties in PN distances are too large to derive confident conclusions.) The results in common for all these galaxies are:

- At a given galactocentric distance,  PN abundances show a much larger dispersion than HII region values. Such a dispersion is larger than the reported uncertainties. 

- A  global galactic enrichment with time is apparent in the sense that central O/H abundances are higher for HII regions than for PNe.

- The  gradients from HIIs are steeper than those from PNe (this is marginal in M\,33). That is, chemical gradients seem to be steepening with time.

But the situation is far from satisfactory, because PN abundances at given galactocentric radii exhibit large scatter, indicating that abundances in PNe are affected by factors still poorly understood (a mixture of populations could be one of them). Besides, even when O/H(PNe)$<$O/H(HIIs) in the center of the analyzed galaxies, at some galactic radius PN and HIIs gradients intercept and, in the periphery the opposite is found: O/H(PNe)$>$O/H(H~IIs), which has no good explanation so far. This occurs in M\,31, M\,81, NGC\,300 and also in our Galaxy (see \citealt{rdi11}), being marginal in M\,33.   Then, possible migration effects and contamination by stellar nucleosynthesis in PNe, or depletion of elements in dust in HIIs, could be artificially flattening the PN gradients in comparison to the HIIs gradients. These and other possible effects have to be analyzed carefully, before extracting conclusions about PN abundance gradients.
 
 \vskip 0.2cm
To improve our knowledge on the subjects mentioned above, it is imperative to analyze 
 the chemistry of larger samples of extragalactic PNe (in particular the faint ones). Obviously, spectroscopy with large-aperture  telescopes (8-10 m) is required (MOS mode is the most efficient technique). Well determined abundance ratios such as He/H, O/H, N/O, Ne/O will help us to better understand the nature of stellar progenitors of extragalactic PNe,  and to constrain better the chemical evolution models of stars and galaxies. Certainly these models need improvements also. A better knowledge of the extragalactic PN populations will be very useful in analyzing the chemical and structural evolution of the parent galaxies.

\subsection{ Kinematics}

The initial motivation for the discovery of extragalactic PNe 
was to use their luminosity function (PNLF) as a distance
indicator (see a recent review by \citealt{ciar12}). Having
detected the PNe, this opened the possibility of using them 
as kinematic tracers in the outskirts of early-type galaxies
\citep{hui95}. PNe were expected to confirm the universality
of dark matter halos in all types of galaxies. 
In recent years the motivation has changed;
now the main interest has become PN kinematics and abundance 
studies, and a PNLF distance is more likely to be a by-product 
of a kinematics-oriented PN search.


\subsubsection{PNe and Dark Matter Halos}
After the classic study of PNe in NGC 5128 (Hui et al. 1995), 
when more efficient methods for PN radial velocity measurements 
were implemented, the new PN data did not always confirm the
presence of dark matter halos around elliptical galaxies. 
To explain this, several arguments were raised. First of all, 
the possible existence of radial anisotropy 
in the velocity distributions, which could mask 
the presence of dark matter (\citealt{metal01}; 
\citealt{dekel05}). But there was also some concern about the 
reliability of the PN radial velocities, and whether or not PNe 
could be trusted as kinematic tracers, because of possible stellar 
population effects.

After more work, this confusing situation has been partially clarified. 
The flattened elliptical galaxy NGC 821 was observed with two different 
telescopes, spectrographs, and slitless velocity measurement 
techniques (\citealt{cocc09}; \citealt{teo10}). The PN 
identifications and radial velocities were in good agreement, and the 
existence of a Keplerian decline of the PN line-of-sight velocity 
dispersion was confirmed.

PN kinematics in a variety of galaxies (including spirals and 
starburst galaxies) have always agreed with other existing tracers 
like HI, CO or H$\alpha$ data, where available for comparison. PNe
are good kinematic tracers of the stellar population they represent
(see \citealt{arna12}).

Finally, consider the massive Virgo elliptical galaxy M 60 \citep{teo11}, where there is independent evidence of dark matter, through 
the presence of hot, X-ray emitting gas. In this case, the PN kinematics 
support the presence of a dark matter halo. The total mass of this halo, 
determined using PNe, is similar to that estimated from globular cluster, 
XMM-Newton, and Chandra data in M 60. Therefore, where independent 
evidence indicates the presence of an extended dark matter halo, PNe 
give compatible results.

\subsubsection{Open Questions and Future Prospects}
What remains for the future is to answer the fundamental question: 
what is the reason for the Keplerian decline of the line-of-sight PN
velocity dispersion in galaxies like NGC 4697 and NGC 821: a comparative 
lack of dark matter, or radial anisotropy? On the one hand, theoretical 
modeling (\citealt{delo08,delo09}) shows that, if we {\it assume}
radial anisotropy, the Keplerian decline can be reproduced in the presence 
of a dark matter halo. On the other hand, it is fair to say that nobody has 
shown definitive evidence that radial anisotropy {\it must}\/ be present. 
One example is NGC 4697 \citep{metal09}. The reader will find 
literature claiming that the best model for NGC 821 requires a dark matter 
halo; but this conclusion was obtained by ignoring the PN data points. 

A way of breaking the degeneracy in this problem is to construct 
histograms of PN radial velocities at different positions across the galaxy.
At large projected distances, the predominance of radial motions should 
produce a peaky distribution; close to the center of the galaxy, there 
should be evidence of a more flattened distribution (in extreme cases, 
a double peak would be expected).

Observationally, this is a difficult problem. To reduce the statistical 
noise, a large number of PNe is required, somewhere between 500 and 1000.
Since PNe follow the light, there will be a low surface density of PNe at 
large projected distances. The solution is to detect fainter PNe, which 
requires a lot of large telescope time, and is complicated by an increasing 
contamination by emission lines from distant galaxies redshifted
into the on-band filter used for PN detection. To illustrate the magnitude 
of the observational problem, consider that the total number of PNe at 
large projected distances discovered in NGC 4697 is 48 \citep{metal09}. We would need to detect somewhere between 10 and 20 times more 
such PNe, distributed across a large field. This is not likely to happen 
before we have 30 meter telescopes.

There is a possible alternative in the direction to the core of the galaxy. 
Such regions have always been rejected as PN searching grounds, because 
of the very high surface brightness, which makes PN detection very hard
if we use the classic onband-offband filter technique. Another complication
is that the detected PN population will be dominated by those close to 
the core, which may not necessarily show as much anisotropy as those 
few which are more distant from the core.

A better way of detecting PNe in a strong background has been recently 
described by \citet{sarz11}. They used the SAURON\footnote{Spectrographic Areal Unit for Research on Optical Nebulae} integral field 
spectrograph. Breaking the anisotropy - dark matter degeneracy will 
require an IFU spectrograph with a larger field of view, attached to
a larger telescope. Such instruments will be available soon, and we 
can hope that this kind of observation will be attempted. 
Whatever the outcome, the resulting information would strongly affect
theoretical efforts to understand galaxy formation. On the dark matter 
side, there has already been some work exploring how a lower dark matter 
content in intermediate ellipticals can be reconciled with the 
predictions from the current cosmological paradigm (\citealt{nap05,nap09}). On the other hand, a solid proof of radial 
anisotropy would be a valuable input for groups trying to simulate the 
formation of elliptical galaxies. What initial conditions dictate the 
presence of substantial radial anisotropy? This is a field where 
observations can guide theory.

We are far from having exhausted the information that can be provided 
by PNe in different kinds of galaxies. There have been studies of PN 
kinematics in one S0 galaxy (NGC 1023, \citealt{noor08}, 
\citealt{cort11}), others are being studied in similar detail.
Studies of PNe in face-on spirals can tell how the velocity dispersion 
perpendicular to the galactic disk changes as a function of distance to 
the center \citep{herciar09}), giving information about the
mass-to-light ratio. Studies of PNe in edge-on spirals like NGC 891
\citep{smen10} can tell how much angular momentum is associated 
with populations located at large distances from the galactic plane; 
which may turn out to be useful in understanding how spiral halos 
are formed.

\subsubsection{The PNe Luminosity function}

In the words of Ciardullo (2012), the PNLF method for distance 
determination cannot work, but it works very well anyway. In other
words, we do not understand, theoretically, why the bright end of 
the PNLF shows its surprising invariance. Clearly this is a problem
that will require future efforts. A solution to this problem can 
potentially teach us a lot about PNe (the old question surfaces 
again: how many PNe are the result of binary star evolution?) and 
their properties in different stellar populations.

Unfortunately, in the near future (before the advent of 30 meter 
telescopes), any satisfactory solution will cost large amounts of 
telescope time. A possible way of improving our knowledge of the
PNLF would be to obtain deep spectrograms of those PNe that populate 
the PNLF bright end. Detection of diagnostic lines like [O III] 
$\lambda$4363 at distances larger than 4 Mpc is too difficult, but 
detection of H$\alpha$ and H$\beta$ would be easier, and would 
provide individual values of extinction and rough limits on the 
oxygen abundance (e.g. \citealt{metal05}, on NGC 4697). It would
then be possible to explore, for example, if internal PN extinction 
plays any important role, or to what extent the distribution 
of intensities of [O III] $\lambda$5007 relative to H$\beta$ should
be modeled more carefully in PNLF simulations.

Another way of testing how well we understand the PNLF is to detect 
fainter PNe, extending our empirical knowledge of the PNLF toward 
its faint end in many galaxies. But again, this will require long 
exposures with large telescopes.

In summary, we can expect a slow increase in the number of PNLF 
distance determinations (which are quite good!) but it is not 
clear if we can expect, in the short term, any sudden progress 
in understanding why they are good. The solution may have to come 
from other areas of PN research.

\end{document}